# Shape-to-Music: A Musical Representation for Structural Topologies of Mechanical Metamaterials

**Sofia Cassara\*, Buminhan Sansa\*, Saltuk Yıldız, Waris Khan, Pınar Acar\*\***

*Department of Mechanical Engineering, Virginia Tech, Blacksburg, VA 24061*

**ABSTRACT**

Mechanical metamaterials are increasingly attracting interest in engineering applications due to their unique mechanical properties and lightweight nature. This study develops a novel sound-based representation to characterize the topologies of mechanical metamaterials, including spinodal designs and porous cellular structures. Two distinct frameworks are introduced: an image-based approach, where material topologies are divided into grids and their relative densities are mapped to evolving melodies, and a numerical simulation approach, where finite element analysis (FEA) visualizes mechanical responses as color-coded images, translated into unique musical compositions. By applying the Fast Fourier Transform (FFT), the generated melodies are analyzed as frequency plots, revealing distinct acoustic signatures for each material topology. This innovative approach not only distinguishes between different metamaterial designs but also provides an intuitive, auditory tool for material characterization. The results demonstrate the potential of sound-based representations to complement traditional methods in materials modeling, offering new avenues for design and analysis.

## 1 Introduction

Mechanical metamaterials provide extraordinary material properties due to their unique structural topologies [1, 2]. These materials have found applications across various industries, including automotive, aerospace, energy, and biomedical sectors, particularly with the recent advancements in fabrication methods [3]. However, conventional metamaterials featuring struts and sharp geometric transitions often suffer from high-stress concentrations, leading to inefficient mechanical responses.

To address these limitations, Kumar et al. [4] introduced a novel class of nature-inspired mechanical metamaterials known as spinodoids. These materials exhibit smoothly varying topological features, drawing inspiration from the thermodynamic phenomenon of spinodal decomposition, in which material phases spontaneously separate [5]. Due to their inherent randomness and geometric complexity, early studies on spinodoids have focused on data-driven approaches for their design and optimization. Zheng et al. [6] utilized deep neural networks to predict the homogenized elastic properties of spinodal metamaterials on a multi-scale level. Similarly, Röding et al. [7] employed convolutional neural networks (CNNs) for the inverse design of these materials with respect to mass transport behavior. More recently, Liu and Acar [8] leveraged conditional generative adversarial networks (CGANs) to perform inverse design of spinodal topologies for target elastic properties. Their approach involved modeling spinodal materials as binary

*\* Equally contributed first authors*
*\*\* Corresponding author (email: pacar@vt.edu)*

images for input into the CGAN framework.

Beyond mechanical applications, spinodal metamaterials have also been explored as acoustic materials. Wojciechowski et al. [9] investigated the sound absorption capabilities of 3D-printed isotropic and anisotropic spinodoid structures using impedance tube testing, further analyzing their bulk properties through the Johnson-Champoux-Allard theory. The non-periodically repeating topology of spinodal metamaterials provides a vast design space, offering new opportunities for structural optimization. Yildiz et al. [10] demonstrated the potential of these materials to exhibit lower stress concentrations compared to conventional metamaterials by employing a deep-learning-based optimization framework.

Porous cellular materials, a common category of nature-inspired engineering structures, also serve as effective mechanical metamaterials for different applications [11-13]. These structures offer significant advantages, such as a high strength-to-weight ratio and the ability to absorb impact energy, making them highly suitable for engineering applications [14-16]. Xue et al. [17] applied neural networks to homogenize the multi-scale nonlinear mechanical behavior of periodically repeating cellular structures. More recently, Liu and Acar [18] explored the material-property spaces of cellular metamaterials by modeling them as representative volume elements (RVEs).

Mechanical metamaterials are increasingly gaining attention for industrial applications, although their potential extends beyond the fields such as automotive, aerospace, energy, and biomedical since they demonstrate exclusive mechanical properties such as low weight and high strength. In this study, we propose and develop the "*Shape-to-Music*" concept as a novel sound-based representation to characterize the unique topologies of mechanical metamaterials. For this purpose, two distinct frameworks are introduced: an image-based approach, where material topologies are divided into grids and their relative densities are mapped to evolving melodies, and a numerical simulation approach, where finite element analysis (FEA) visualizes mechanical responses as color-coded images translated into unique musical compositions. Both approaches are tested using the image data of spinodal topologies and porous cellular structures. The findings outline that the proposed approaches can detect the differences between the geometries of different topologies and the periodicity of cellular designs. "*Shape-to-Music*" is developed as an interactive learning approach that helps users explore materials design and mechanical metamaterial topologies by identifying similarities and differences through musical representations.

The paper is organized as follows. Section 2 details the computational design of spinodal and porous cellular metamaterials. Section 3 explores two music generation approaches, namely image-based and FEA-based methods. Section 4 presents the results, and Section 5 concludes the study with key findings.

## 2 Computational Design of Mechanical Metamaterials

This section discusses the mathematical modeling and quantification of the topologies of two distinct classes of mechanical metamaterials: (i) spinodal topologies and (ii) porous cellular structures.

### 2.1 Modeling Spinodal Topologies

The *Shape-to-Music* concept is first applied to spinodal topologies that are numerically generated in 2D as binary images composed of solid and void phases. The phase-fields ($\psi(x)$) of the spinodal topologies are computed using Gaussian Random Fields (GRFs), and the final phases are determined based on a level-set approach as given in Eq. 1 [4].

$$\psi(x) = \sqrt{\frac{2}{N}} \sum_{i=1}^{N} \cos(\beta n_i \cdot x + \gamma_i) \tag{1}$$

The material phase of the spinodoid unit cell is described by $N$ number of cosine waves, and $\beta$ stands for the length scale, representing the distance between two consecutive occurrences of the same phase, the orientation of distributions ($n_i$), and the phase angle ($\gamma_i$).

$$n_i \sim U(\{k \in S^2 : (|k.\hat{e}_1| > \cos(\theta_1)) \oplus (|k.\hat{e}_2| > \cos(\theta_2))\}) \tag{2}$$

In Eq. 2, the randomized orientation distributions determine the final spinodal shapes. In the computational design framework, the cone angles ($\theta_1$, $\theta_2$) are restricted to be between $15^o$ and $90^o$. Furthermore, the number of cosine waves ($N$) and constant wave number ($\beta$) are selected as 3 and $6\pi$, respectively. The threshold porosity is kept at 0.35, and the volume fraction ($v_f$) of each geometry is around 60%.

The solid and void phases are mathematically demonstrated with a binary indicator as follows [19]:

$$X(x) = if\ \psi(x) < \emptyset_0\ = 1\ ; else\ = 0 \tag{3}$$

where $\emptyset_0 = \sqrt{2} erf^{-1}(2\rho_{rel} - 1)$ [4] is the function to calculate threshold value ($\emptyset_0$) related to the relative density ($\rho_{rel}$). As given in Eq. (3), the material fields satisfying the level-set (i.e., $\psi(x) < \emptyset_0$) are retained as the solid phase (e.g., black), while the remaining material defines the void phase (e.g., white). Exemplary binary images of spinodal topologies generated with the presented formulation are demonstrated in Fig. 1.

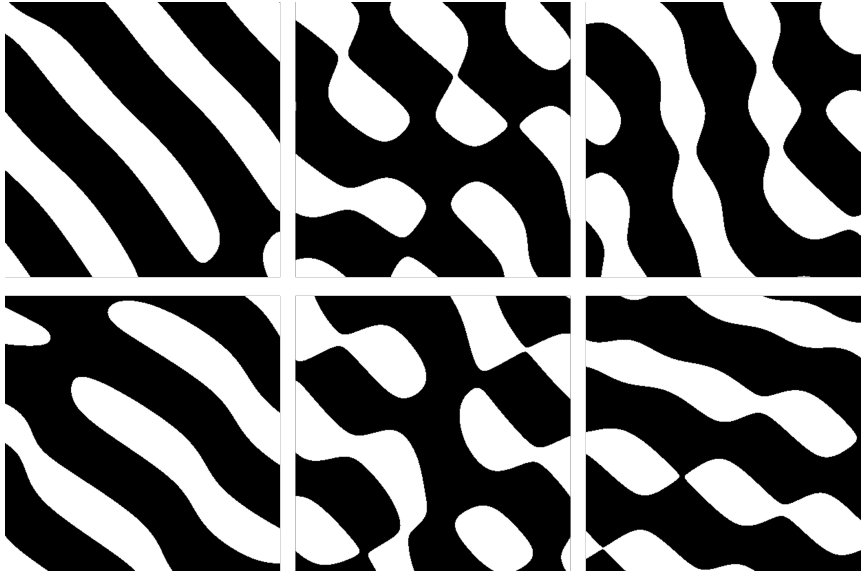

Figure 1: Example spinodal topologies generated with the GRF-based formulation.

### 2.2 Modeling Porous Cellular Structures

The *Shape-to-Music* concept is also applied to 2D porous cellular topologies that are generated using the analytical expressions shown in Eq. 4 and Eq. 5. The features of porous cellular structures are mainly based on the topology itself, such as the shape of the pore, as shown in Fig. 2. The unit cell topologies of the porous cellular structures are defined by Eq. 4 [17]:

$$r(\theta) = r_0(1 + \zeta_1 \cos(m.\theta) + \zeta_2 \cos(n.\theta)) \tag{4}$$

$$r_0 = L\sqrt{\frac{2\varphi}{\pi(\zeta_1^2+\zeta_2^2+2)}} \tag{5}$$

In Eq. 4, $\theta$ is the power angle swept between the range of $0 \leq \theta \leq 2\pi$, $r$ stands for the polar radius, L is the cell length, and $\varphi$ shows the cellular structure porosity. The coefficients, $n$ and $m$, along with other design parameters, $\zeta_1$ and $\zeta_2$, determine the resulting cellular structure topology. Figure 2 demonstrates several sample geometric patterns of unit cells with various pore shapes controlled by the parameters $\zeta$. These porous cellular structures are generated using fixed values of $L$ and $\varphi$ equal to 10 mm and 0.45, respectively.

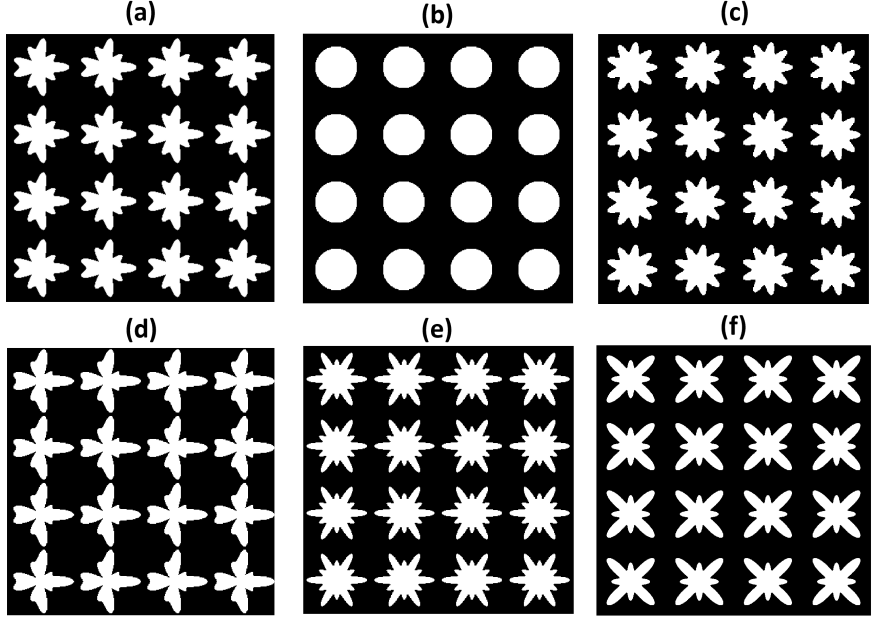

Figure 2: Sample porous material properties generated with the following parameters: (a) $\zeta_1 = 0.25$, $\zeta_2 = 0.25$, $m = 4$, $n = 9$; (b) $\zeta_1 = 0$, $\zeta_2 = 0$, $m = 4$, $n = 9$; (c) $\zeta_1 = 0$, $\zeta_2 = 0.25$, $m = 4$, $n = 9$; (d) $\zeta_1 = 0.5$, $\zeta_2 = 0.25$, $m = 4$, $n = 9$; (e) $\zeta_1 = 0.25$, $\zeta_2 = 0.25$, $m = 6$, $n = 12$; (f) $\zeta_1 = 0.35$, $\zeta_2 = -0.35$, $m = 8$, $n = 4$.

## 3 Introduction of *Shape-to-Music*

The *Shape-to-Music* concept is tested with spinodal and porous cellular structures. The goal of this concept is to produce music that reflects material similarities such that similar structural topologies generate similar melodies, while distinct topologies yield unique tunes. As demonstrated in Fig. 3, the spinodal structures are anticipated to produce music involving non-periodically repeating melodies as a result of their non-periodic topologies. On the other hand, the porous cellular structures are expected to generate periodically repeating tunes, reflecting the periodic nature of their topologies. The tunes are created by following two different numerical methods, image-based and FEA-based approaches, as explained next.

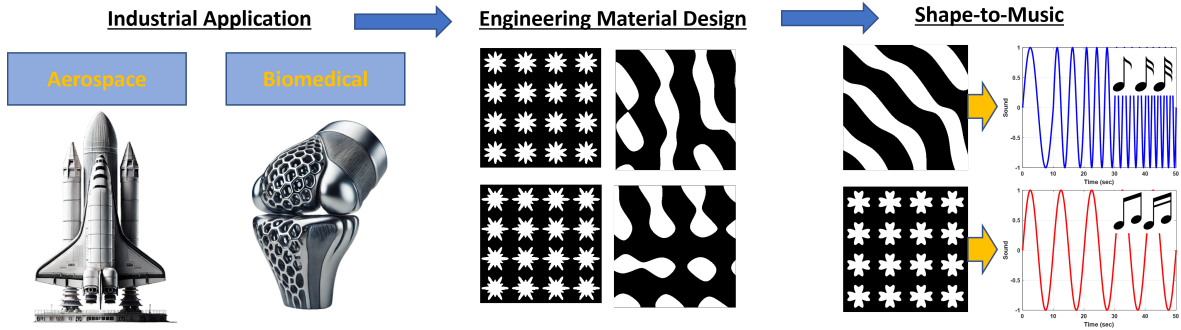

Figure 3: Schematic representation for the underlying goal behind the *Shape-to-Music* concept, which is to develop a novel sound-based representation to characterize the topologies of mechanical metamaterials. The proposed methods are applied to spinodal topologies and porous cellular materials, which are candidate structures for aerospace and biomedical industries due to their unique mechanical properties and lightweight nature.

## 3.1 Image-Based Approach

The image-based approach transforms the metamaterial images into music using the image's intrinsic characteristics, specifically the ratio of black pixels to total pixels. This method allows for straightforward song creation without relying on external criteria, making the unique spinodal topology patterns the sole independent variable. This approach is developed using MATLAB. There are two associated MATLAB functions created to generate music: S2M and S2M_compare. S2M is used to create a song for a single image, while S2M_compare analyzes the songs (made by calling S2M) for two different images.

In calling S2M, the user selects the (metamaterial) PNG image to generate a song for. The image is processed in MATLAB and resized to be $1000 \times 1000$ pixels. The grayscale images can be characterized by such two-dimensional matrices. The grayscale image is updated to be an RGB (red-green-blue) image by resizing it to be a three-dimensional array. The significance of having a "three-dimensional" color representation will be demonstrated later when the image is color-blocked and displayed back to the user. The program then divides the image into sub-images, allowing the user to choose the size of the sub-image (length) from 50, 100, 200, 250, and 500 pixels. This side length is directly related to how close the processing method follows the original image, and is inversely related to the length of the song. The recommended box length is 100 pixels, as this length reflects changes within the original image while yielding a song approximately 50 seconds in length. The process is visualized in Fig. 4.

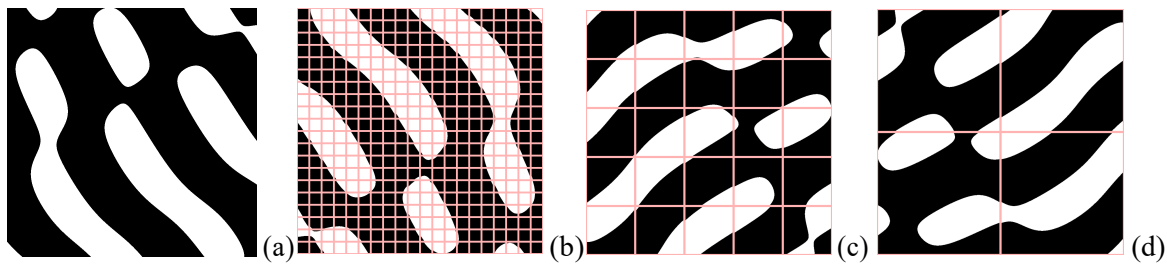

Figure 4: (a) Initial Spinodoid (b) with $50 \times 50$ pixels sub-images (b) with $200 \times 200$ pixels sub-images (d) with $500 \times 500$ pixels sub-images

At each corner sub-image, the black-to-total pixel ratio, which will be referred to as "density", is calculated. The densities at each corner are compared, and the entire image rotates so the densest corner is in the top left position (if it is not already in this position) (as shown in Fig. 5). This idea of density-dependent rotation acts as a standardization of the images, which serves to counteract anisotropy. Once this corner rotation is completed, the program calculates the density within each sub-image, beginning at the top left corner and ending at the bottom right corner. These densities are stored in a matrix, where the location of each entry matches the location of its corresponding sub-image.

A set of note frequencies (ranging from C3 to C7 on a piano) and colors is established to correspond to density ranges of 10%; a density of 0% (no black pixels present) is assigned the lightest color and largest note frequency (which produces the highest note), while a density of 100% (all black pixels) is assigned the deepest color and smallest note frequency (which produces the deepest note). The program iterates through the density matrix, assigning each sub-image its respective note frequency and color. The frequencies are stored in a matrix, while the colors fill their corresponding sub-image. The program then displays the original image (post-rotation) and the color-blocked image side-by-side, which allows the user to visualize density trends across the image.

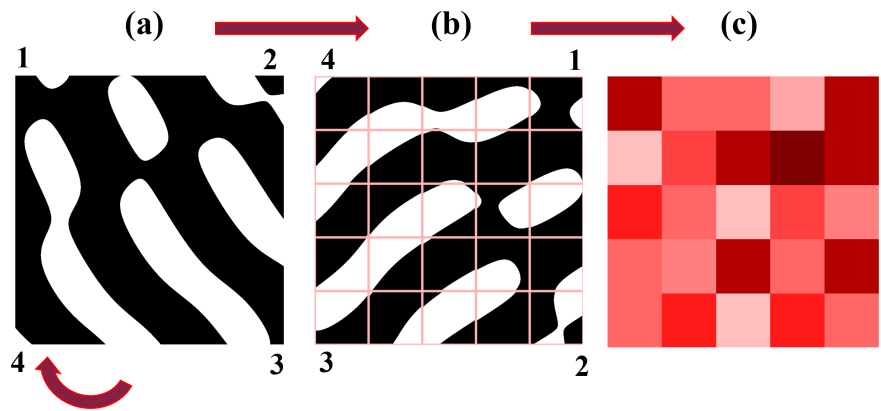

Figure 5: (a) Original Spinodoid (b) Spinodoid post-rotation (c) Color-Blocked Image

The song is created by iterating through the matrix of note frequencies and multiplying the frequency within a sine function (shown below) to create the corresponding note. The time used in the equation is a vector with a number of elements corresponding to the sampling frequency, ranging from 0 to the note duration. The MATLAB sound function is then used to play the note. The program gives the user a choice of playing the song aloud or not; if not, the sound function is not used, but the rest of the process remains identical. The program also gives the user a choice of saving the song; if yes, the song is saved as a ".wav" file, with the original image's name and sub-image length used as the song title.

$$note = \sin(2\pi . \omega_{note} . t) \tag{6}$$

After the song has been created, the program generates a list of "statistics" for the image consisting of sub-image length, average density per sub-image, average note, average color (in RGB format), mode note, and mode color (in RGB format). The average and mode colors are also indicated on a gradient diagram composed of the previously established color scale (Fig. 6).

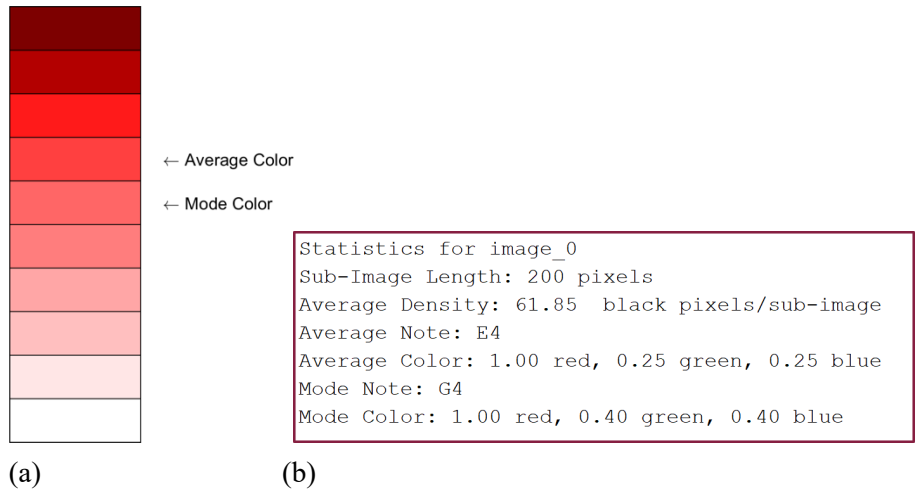

(a)             (b)

Figure 6: Statistical Outputs: (a) Color Gradient (b) Command Window Statistics

The function S2M_compare prompts the user for the names of two images to compare songs for, then calls S2M for each image. The original image and color-blocked image for both inputs are displayed in one figure (as in Fig. 7), allowing the user to compare density trends across both images. For every statistic mentioned above, S2M_compare evaluates the similarities and differences between the images. Comparing the songs requires both images to have the same sub-image length, and is done by comparing the note frequencies at each sub-image. For each frequency difference between the two images, the script displays the location, density, and note for the corresponding sub-image. An example output has been sampled in the charts shown in Fig. 8.

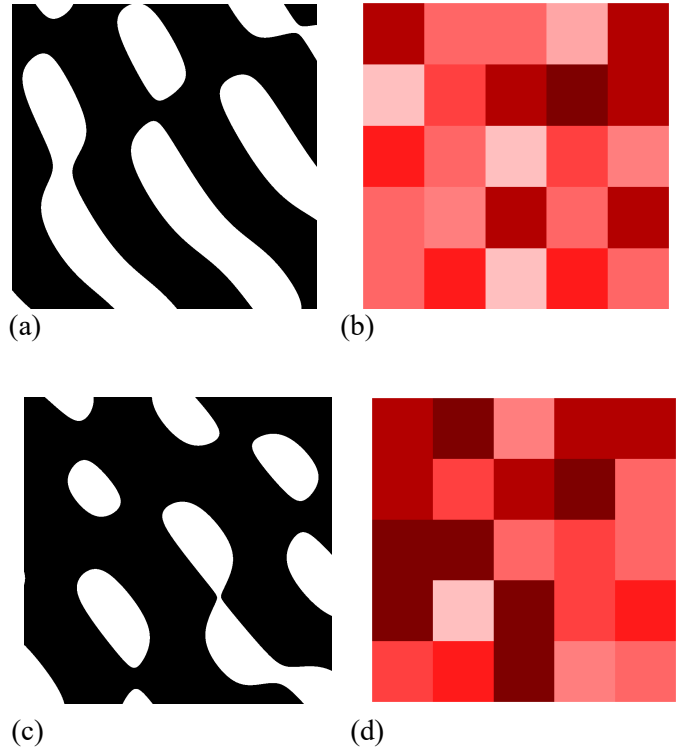

Figure 7: Image Comparison: (a) Original Image 0 (b) Color-Blocked Image 0 (c) Original Image 1 (d) Color-Blocked Image 1.

| Sub-Image (1:200, 201:400) | Image 0 | Image 1 |
|---|---|---|
| Sub-Image Length | 200 pixels | 200 pixels |
| Density (pixels) | 54.40 | 96.68 |
| Note | G4 | C3 |
| | Image 0 | Image 1 |
| Average Density (pixels/sub-image) | 61.85 | 73.58 |
| Average Color | 1.00 R, 0.25 G, 0.25 B | 1.00 R, 0.10 G, 0.10 B |
| Mode Color | 1.00 R, 0.40 G, 0.40 B | 0.49 R, 0.00 G, 0.00 B |
| Average Note | E4 | E4 |
| Mode Note | G4 | C3 |

Figure 8: Comparison Outputs

## 3.2 Finite Element Analysis (FEA)-Based Approach

The FEA-based approach focuses on structural analyses of metamaterial topologies. This approach incorporates new variables that critically affect the sound output of each metamaterial. Key factors include geometric parameters (surface roughness, porosity), types of loading (tension, compression, shear), boundary conditions, the specific locations where the stresses are applied, and the resulting displacement fields under these conditions. Figure 9 shows the simulation setup (i.e., assignment of boundary conditions and applied loads, as well as material phases) for FEA of a randomly generated spinodal topology subjected to non-dimensional tensile loads with fixed boundary conditions.

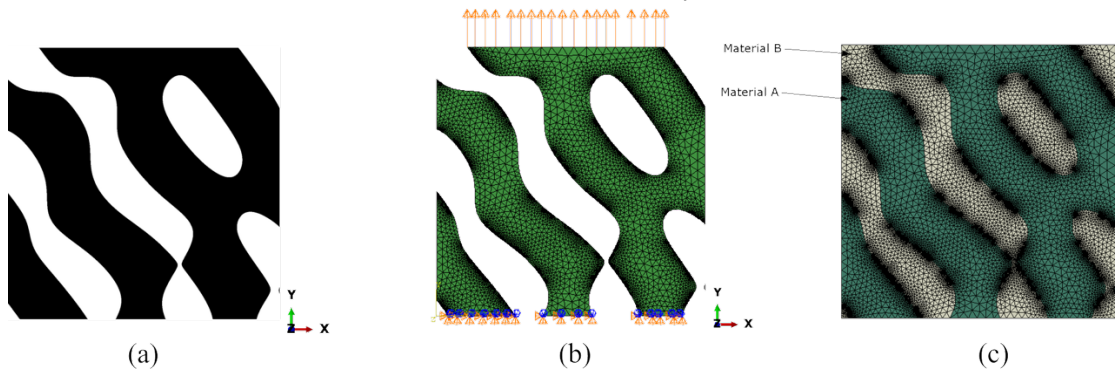

(a)         (b)         (c)

Figure 9: The FEA-based framework. **(a)** Spinodal topology; **(b)** the visualization of boundary conditions and applied loads; **(c)** the assignment of phases and the finite element mesh. Materials A and B correspond to solid and void phases of the spinodal topology, respectively.

The steps outlined below are followed to create the music representations for mechanical metamaterials using the FEA-based approach. The 5-step procedure is explained for an example spinodoid, while the method is applied to both spinodal topologies and cellular metamaterial designs in Section 5.

*Step 1) Creating FEA Mesh*

The mechanical metamaterial geometries generated using the formulation in Section 2 are then processed in MATLAB, where the 2D topologies are converted into a finite element mesh using an online tool, *Im2Mesh* [20]. Meshing is performed using *Im2Mesh* due to its capability to generate finite element meshes automatically for the binary images of mechanical metamaterials. The meshing elements generated by *Im2Mesh* consist of two-dimensional triangular elements. A higher-order mesh is not applied in this analysis, as the image is modeled as a two-dimensional structure. Due to the linear interpolation of displacement in first-order triangular elements, the strain within each element is constant. Since stress is directly derived from strain, it also remains constant within the element. This behavior is a consequence of the linear interpolation used in first-order elements, which results in coarse strain and stress contour plots when displacement is differentiated. As displacement is interpolated across nodes, displacement contour plots show a smoother transition compared to a stress contour plot. However, employing higher-order elements would allow for variable stress and strain distributions within an element, improving the accuracy of stress contour plots. The use of higher-order elements is not necessary for this analysis, as the current mesh resolution is deemed sufficient.

## *Step 2) Transferring FEA Mesh to FEA Software*

The FEA is performed using the ABAQUS software. A Python script is developed to transfer the finite element mesh created in *Im2Mesh* in Matlab to ABAQUS. The script imports the necessary modules to set up and execute the FEA task. The script assigns two materials, Material A for the solid (black) regions and Material B for the void (white) regions of the topology.

The solid phase (Material A) is modeled as the Ti-6Al-4V alloy, with a modulus of elasticity ($E$) set to 110 GPa and a Poisson's ratio of 0.3 [21, 22]. Material B, representing the void regions, is assigned to an ersatz material phase to ensure the stability of FEA, and thus its modulus of elasticity is set to an infinitesimal value (1$e$ 15), and the Poisson's ratio is set to 0. This infinitesimal modulus is critical to define Material B as effectively non-load-bearing, ensuring it does not contribute to the structural analysis in the ABAQUS FEA simulation for Material A. The script continues to create a static analysis step, applying fixed boundary conditions at the bottom of the metamaterial geometry ($u_1 = 0$, $u_2 = 0$), while a tensile force is applied at the top, and enabling vertical displacement (Figure 9 (b)). Finally, the script generates and executes a job in ABAQUS for each metamaterial topology, iterating over the provided set of images.

## *Step 3) FEA using ABAQUS*

In this step, the FEA simulation is executed using ABAQUS. Once the script execution is complete, the results for the total displacement magnitude (Figure 10 (a)) and von Mises stress (Figure 10 (b)) are obtained and visualized.

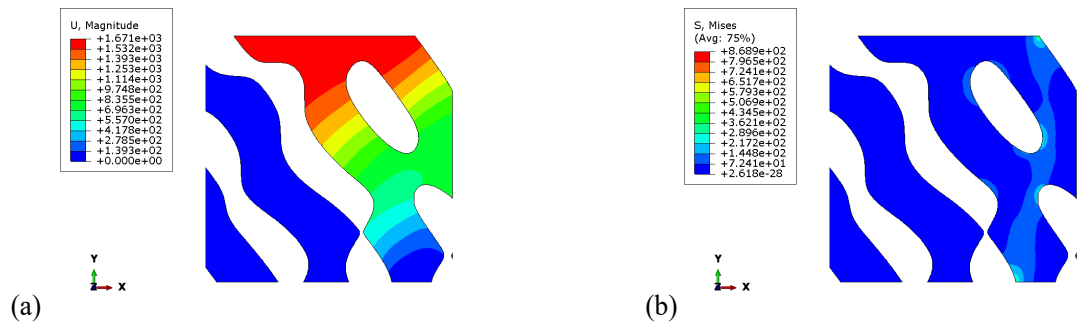

(a)          (b)

Figure 10: FEA simulation results showing the distributions of (a) total displacement magnitude and (b) von Mises stress. This figure is used to detect the ranges of color variation in displacement and stress fields. The displacement distribution provided in Figure (a) yields a greater distribution of colors and, thus, frequencies.

The displacement magnitude is selected as the prime parameter to obtain color data due to its broader color intervals compared to other static outputs (von Mises stress). This wider range of colors grants a more distinguishable sound differential when converting the colors to frequencies. Post-processing the output image from ABAQUS ensures the results are suitable for MATLAB to compute frequencies with maximum accuracy. The post-processing steps include setting the deformation scale factor to 0, changing the *exterior edges* display to *free edges*, setting the render style to *filled*, and disabling all viewport annotation options. These adjustments are essential for ensuring that the MATLAB *Shape-to-Music* code

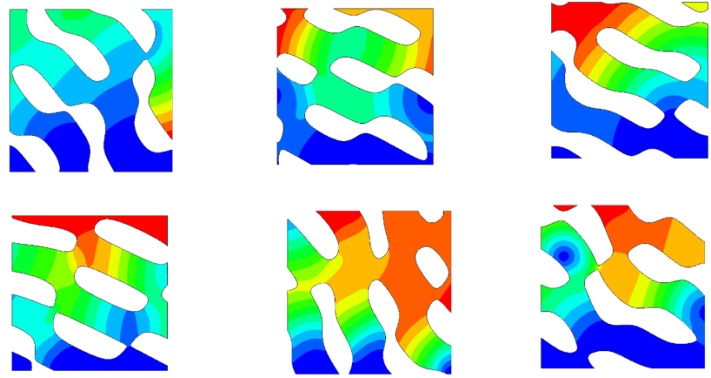

Figure 11: Color maps obtained for example mechanical metamaterial topologies using the *Shape-to-Music* code.

### *Step 4) Generation of Music from Color Maps*

In this step, the MATLAB script transforms displacement contour maps from ABAQUS into unique musical compositions. First, the image is loaded using MATLAB's *imread* function and resized from the original to allow for a shorter song duration. An independent step-size parameter is developed to control the duration of the song. A larger step size means that *Shape-to-Music* calculates the frequencies every large step, missing pixels in between that can lead to a distinct sound difference between two different topologies. Based on several trials, a step size of 5-7 is recommended. A sampling frequency of 10.1 kHz is then set for clear audio and a base duration for each note of 0.15 seconds. *Shape-to-Music* then assigns distinct frequency ranges to the three RGB channels of the image. The lower and upper bound values are set to 50 and 300 Hz for the red channel, 301 and 1000 Hz for the green channel, and 1001 and 4500 Hz for the blue channel, respectively. These ranges represent different musical ranges, with red capturing bass and sub-bass tones, green showing lower mid-range tones, and blue corresponding to high mid-range and presence tones, respectively. For each pixel in the image, normalized RGB values are calculated by scaling the original color intensity (0-255) to a range of 0-1. The normalized color is mapped back to its respective frequency range, and the average of the transformed frequencies is computed to represent the total frequency output of the pixel (Eq. 7). The loop uses a step size of 6 pixels when reading the image to reduce the density of the sound and length of the audio. This approach is critical for capturing the unique audio output each metamaterial generates. Each material has a unique response to the static analysis, resulting in a different contour plot for each topology.

$$R_{Freq} = R_{Min} + (R_{Max} - R_{Min}) \times R_{normalized} \qquad (7)$$

*Shape-to-Music* uses these transformed frequency values to generate sine waves. A sine wave is created for each frequency, representing the sound associated with a specific step size of the pixel's color. The duration of each sine wave, given by Eq. 6, is fixed at 0.15 seconds to ensure uniform note lengths. *Shape-to-Music* also creates a histogram of the frequency for a clear understanding of the frequency differential, and which frequency dominates the music. For almost all tested spinodal topologies, 450.67 Hz is found to be the dominating frequency. This is because MATLAB reads the white pixels as (*R, G, B*) = (0*, 0, 0*). When a void is introduced, this leads to the same calculation, resulting in a dominating frequency at a fixed value. *Shape-to-Music* can be changed to skip the void areas and calculate a frequency of 0 Hz, resulting in no sound. This is not implemented because the void areas of the mechanical metamaterials are its attributes, as they need to be considered for the continuity of the sound representation, and they are also assigned a secondary material in FEA.

After all pixels are processed according to the step size, the resulting sine waves are combined into a unique audio representing the topology. This method offers a unique and innovative way to translate the visual data of mechanical metamaterial topologies into music.

## 4 Virtual Interface Development

The Shape-to-Music project (accessible at https://astro-lab-vt.github.io/shapetomusic/) provides an interactive visual interface that serves as a repository for music generated through two distinct methods: Image-Based and Finite Element Analysis (FEA)-based approaches, as shown in Fig. 12. The main page introduces the project's objectives and directs users to dedicated interfaces for each method.

The Image-Based Approach presents a selection of cellular and spinodal metamaterial images as shown in Fig. 13. Selecting an image leads users to its dedicated sub-page, which serves as a repository for its musical representations. Here, users can visualize and hear how varying sub-image sizes influence the resulting music. These sub-pages allow users to explore how spatial patterns and geometric features are translated into music. Fig. 14 shows one of the sub-pages when the user clicks on the image CMM3 of the screen, which is shown in Fig. 13.

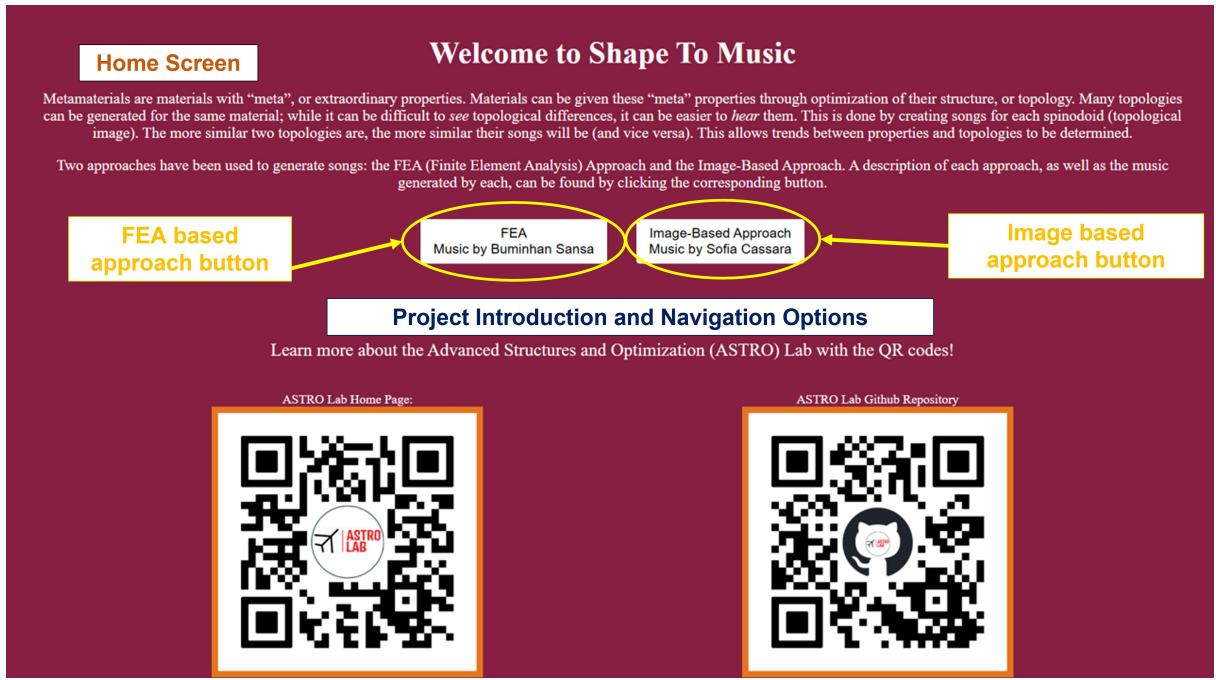

Figure 12: The annotated home screen of the developed interface (https://astro-lab-vt.github.io/shapetomusic/).

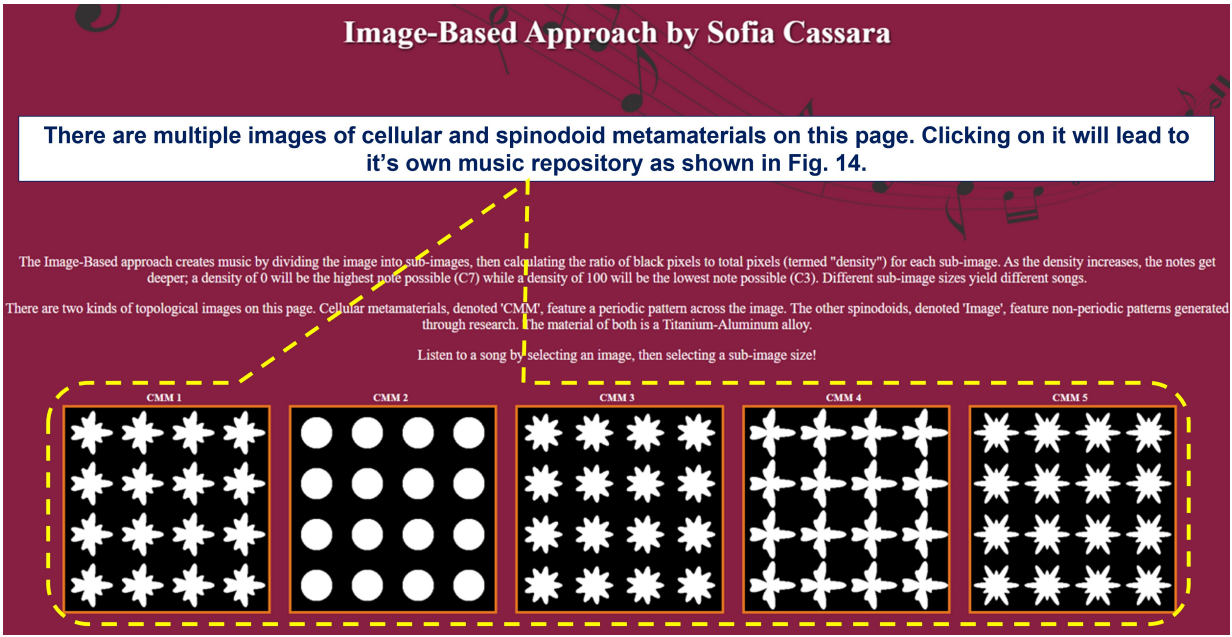

Figure 13: The window which opens up when clicked on the "Image-based approach" button of the home screen in Fig. 12. The figure is annotated to provide instructions.

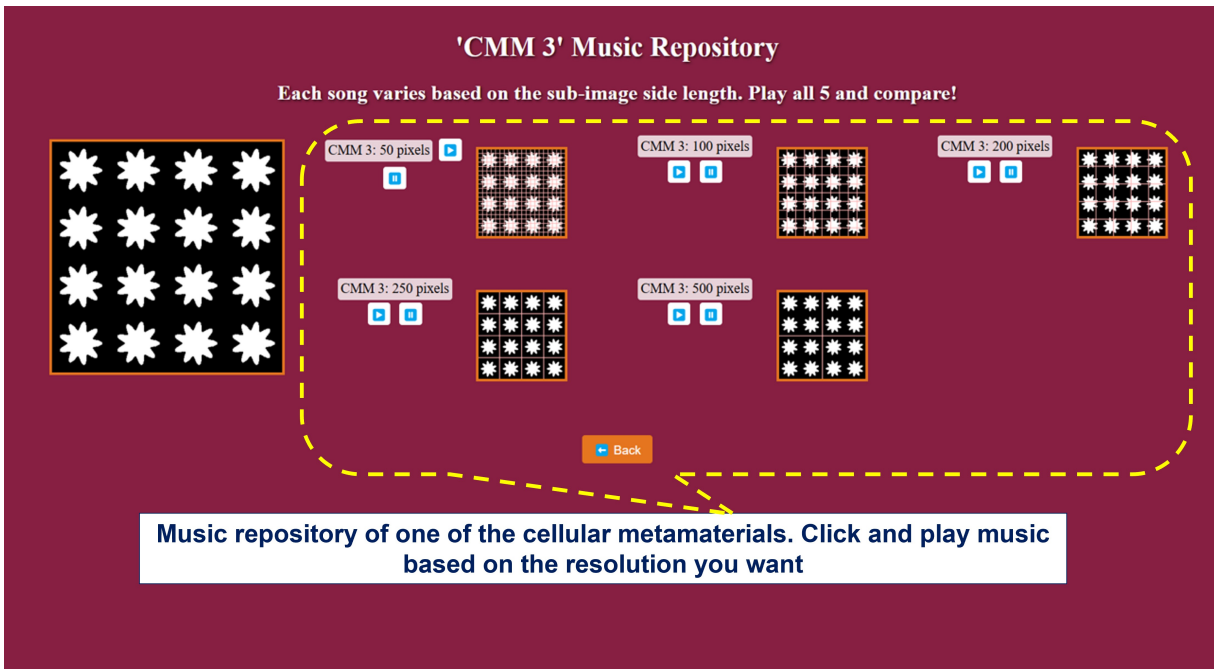

Figure 14: The sub-page when the user clicks on the image CMM3 of the screen which is shown in Fig. 13. The figure is annotated to provide instructions.

In contrast, the FEA-Based Approach showcases a distinct set of spinodal metamaterials, pre-analyzed using finite element methods. Interactive displacement contours illustrate each structure's simulated mechanical response. Selecting an image initiates playback of the sonified FEA data, effectively converting structural mechanics, such as deformation and stress fields, into music. Fig. 15 is the window that opens up when the user clicks on "FEA-based approach" at the home screen in Fig. 12. Thus, the platform provides open access to the project's results and helps users understand how material shapes and mechanical properties can be represented as music.

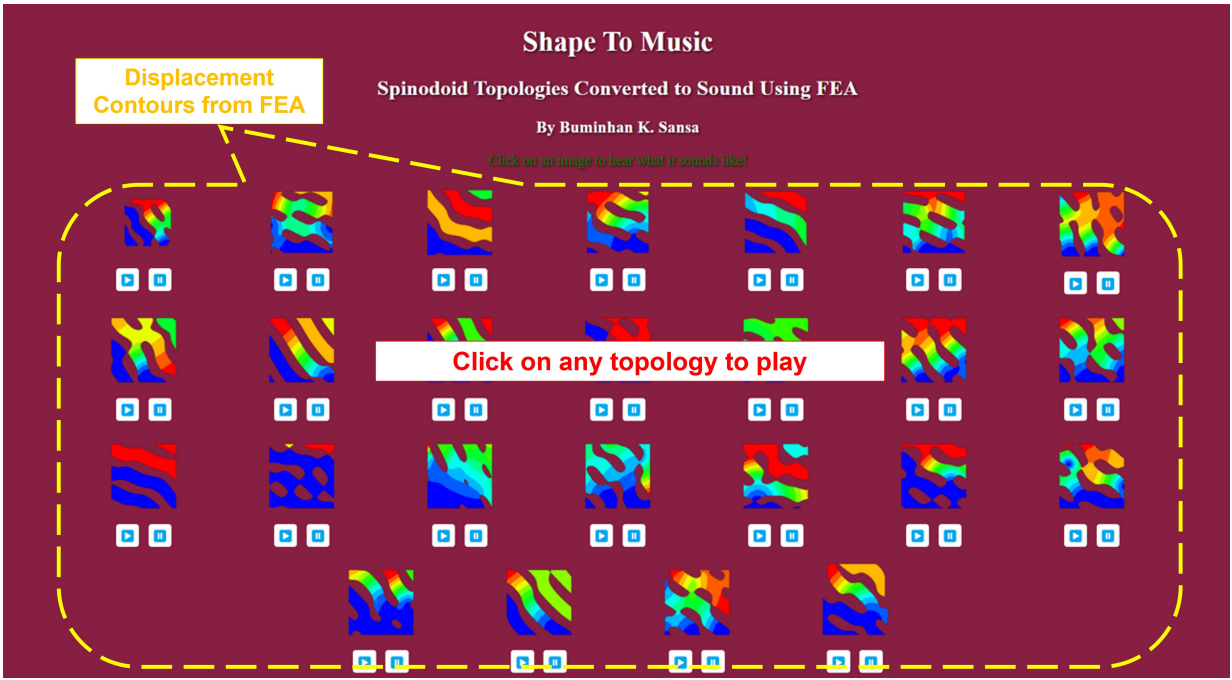

Figure 15: The window which opens up when clicked on the "FEA based approach" button of the home screen in Fig. 12. The figure is annotated to provide instructions.

## 5 Results and Discussion

This section presents frequency plot results for spinodal topologies and porous cellular mechanical metamaterials using image-based and FEA-based approaches.

The image-based method yields objective material analysis through graphical frequency comparison (Fig. 16 (a)-(d)). Frequency has a direct relationship with the angular frequency of the wave (see Eq. 6); as the frequency associated with a sub-image increases, the angular frequency (rate of change of the wave phase) increases. This means that the wave angle achieves a full circle at a faster rate, decreasing the wave's period (time to complete one oscillation). This translates to the graph as striations across the figure. Larger frequencies (higher notes) have more distinct amplitudes than smaller frequencies (lower notes); they appear 'needle-like' toward the amplitude's extremes and appear more concentrated around the $x$-axis. In contrast, smaller frequencies appear concentrated throughout their entire band, demonstrating less of this needle effect. Frequencies in the middle of the range appear as 'light bands' in the graph; this can be attributed to them exhibiting behavior of both low and high frequencies. This data is also represented in the Fast Fourier Transform (FFT) graphs, where the magnitude of the frequencies for each image is depicted in the frequency domain. The spinodoids exhibit a wide range of frequencies, with the largest magnitudes occurring between 200 and 400 Hz. In contrast, the cellular metamaterials exhibit a narrower range of frequencies, and the largest frequency magnitude occurs between 300 and 400 Hz.

For the FEA-based approach, the sine wave and FFT spectrum plots (Fig. 16 (e)-(h)) provide insight into how spinodoid displacement contour maps are converted into sound. The sine wave representation plot shows the time-domain waveform, and each oscillation corresponds to a specific frequency from the color-mapped displacement images. The *x*-axis represents time (seconds), plotting through the whole song duration. The *y*-axis represents amplitude, showing how the sine wave changes over time. The plot itself is a combination of multiple sine waves, each wave corresponding to the step-size pixel frequency from the color-mapped images. The plot shows a high density of oscillations throughout, indicating a wide distribution of frequencies present in the spinodoid music. Lower frequencies appear as smoother oscillations, while higher frequencies oscillate much faster, creating sharp peaks. This explains the "needle-like" effect shown towards the extremes, and the dense waveform in the middle is due to overlapping mid-range frequencies creating a superposition effect.

Meanwhile, the FFT spectrum plot converts the time-domain waveform into the frequency domain. The FFT spectrum plot identifies whether the generated music is dominated by low, mid-range, or high frequencies. The *x*-axis represents frequency in Hertz, ranging from 0 to the Nyquist frequency (Sampling Frequency/2) to cut out duplicate frequencies. The *y*-axis represents magnitude, revealing the most dominant frequencies of the spinodoid. The spinodoids exhibit a wide range of frequencies, with the largest magnitude, around $10^4$, occurring just before 500 Hz, followed by a range of peaks with a $2\times10^4$ magnitude between 750~1750 Hz. The cellular metamaterials exhibit a larger range of frequencies, with the largest magnitude, around $7\times10^4$, occurring just before 500 Hz, followed by a range of peaks with a $3\times10^4$ magnitude between 750-2000 Hz.

The symmetric pattern depicted in the cellular metamaterial sine wave representation plots and the narrow range of frequencies within their FFT spectrum plots reflect their periodic structure. In comparison, the scattered distribution of frequencies throughout the spinodoid figures reflects their non-uniform, optimized spatial structure.

# 6 Conclusion

This study introduces the *Shape-to-Music* concept as a novel approach to characterize the topologies of mechanical metamaterials through sound-based representations. By implementing image-based and finite element analysis-based frameworks, the geometric features of spinodal and porous cellular metamaterials are successfully translated into musical compositions. The findings indicate that different topologies produce distinct auditory patterns, enabling the differentiation of structures based on their unique geometric properties. Furthermore, the periodic nature of cellular metamaterials is observed in both the frequency plots and the generated sound representations, highlighting the relationship between geometric regularity and musical periodicity. This interdisciplinary approach provides a new perspective on material characterization and offers potential applications in design optimization, material identification, and creative exploration in engineering and the arts.

**Disclosure Statement**

No potential conflict of interest was reported by the authors.

**Data availability**

The database for the *Shape-to-Music* website is shared in the ASTRO Lab GitHub repository:
https://github.com/Astro-Lab-VT/shapetomusic

**Funding**

This research was supported by the National Science Foundation (NSF) CAREER Award CMMI-2236947.

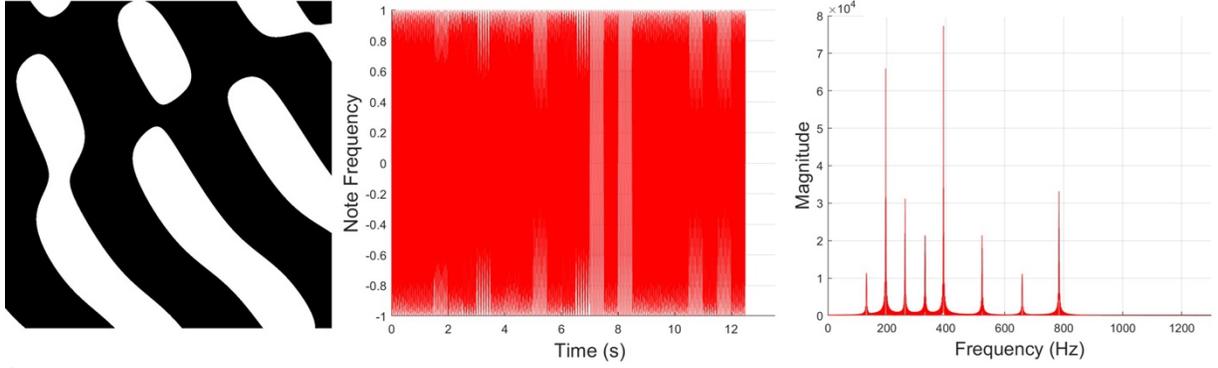

(a)

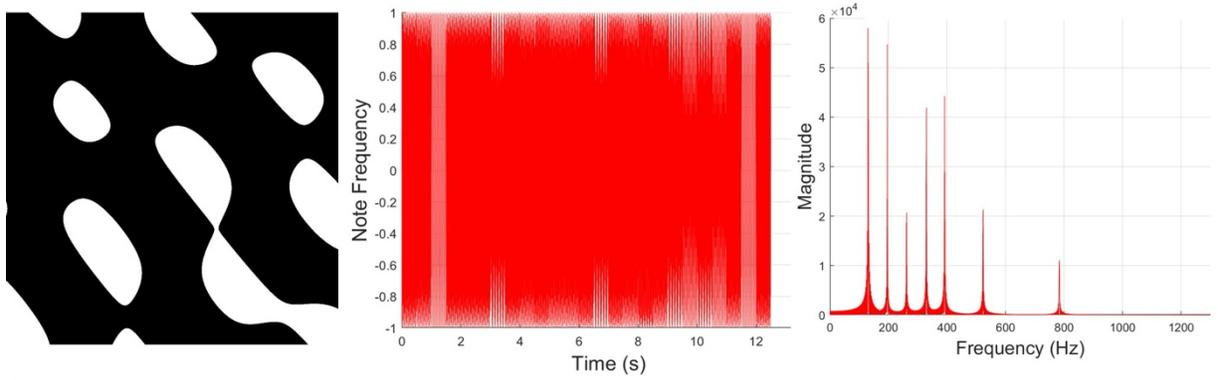

(b)

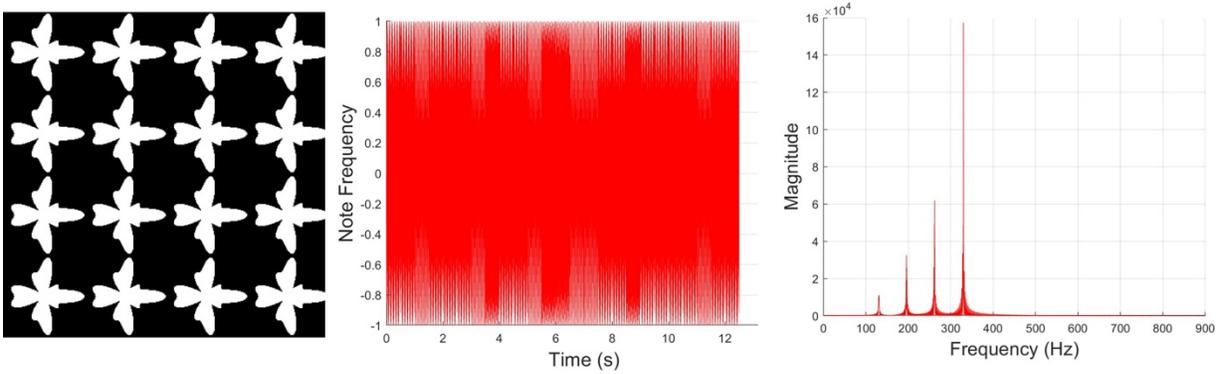

(c)

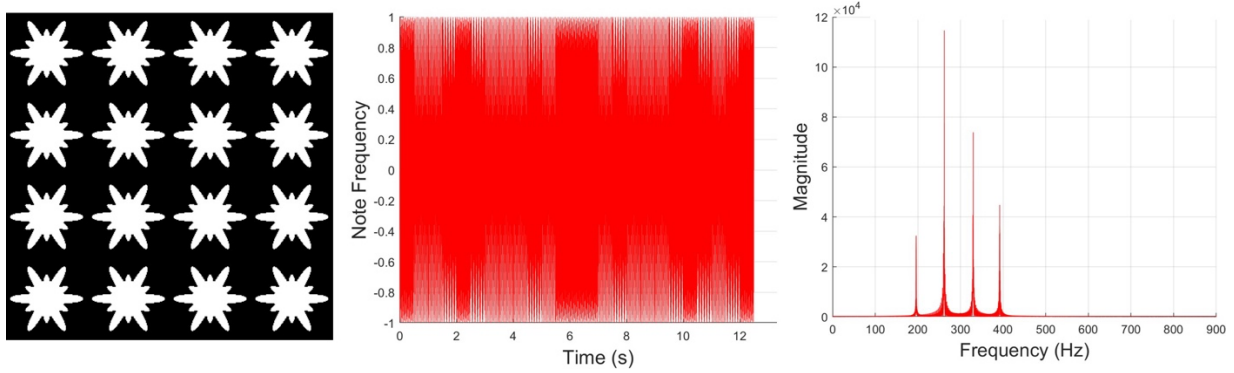

(d)

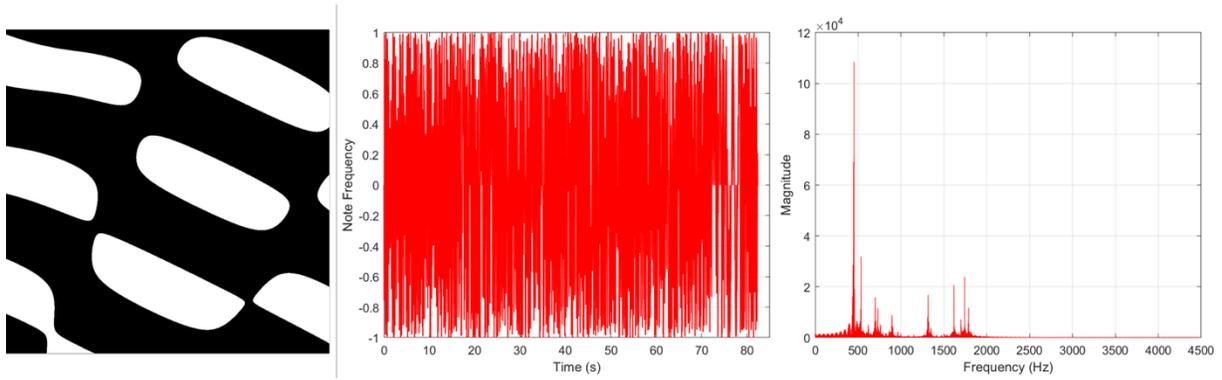

(e)

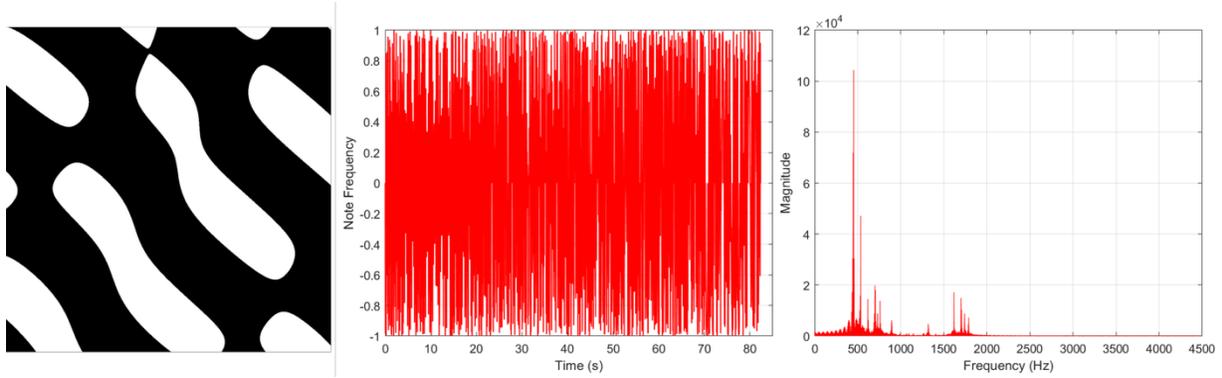

(f)

Figure 16: Frequency comparisons for different mechanical metamaterial topologies using the Image-Based ((a)-(d)) and FEA-based ((e)-(h)) approaches.